\documentclass[useAMS]{mn2e}
\usepackage{graphicx,amssymb}
\title[Cosmic Ray Origins in Supernova Blast Waves]
{Cosmic Ray Origins in Supernova Blast Waves}
\author[A.R. Bell]
{A. R. Bell\thanks{E-mail:t.bell1@physics.ox.ac.uk}\\
Clarendon Laboratory, University of Oxford, Parks Road, Oxford OX1 3PU, UK\\ }

\begin{document}
\date{}
\pagerange{\pageref{firstpage}--\pageref{lastpage}} \pubyear{2014}
\maketitle
\label{firstpage}
\begin{abstract}
We extend the self-similar solution derived by 
Chevalier (1983a) for a Sedov blast wave accelerating cosmic rays (CR)
to show that the Galactic CR population can be divided into:
(A) CR with energies above $\sim 200$GeV released upstream during CR acceleration by 
supernova remnants (SNR), 
(B) CR advected into the interior of the SNR during expansion and then 
released from the SNR at the end of its life to provide the Galactic CR component below $\sim 200$GeV.
The intersection between the two populations may correspond to a measured change in the 
Galactic CR spectral index at this energy (Adriani et al 2011).
\end{abstract}

\section{Introduction}

Supernova remnants (SNR) are the most probable source of Galactic cosmic rays (CR)
at energies up to a few PeV.
CR gain energy at the outer shocks of supernova blast waves by 
first order Fermi diffusive shock acceleration 
(Krymsky 1977, Axford Leer \& Skadron 1977, Bell 1978, Blandford \& Ostriker 1978),
although second order Fermi processes may also contribute (Ostrowski 1999).
CR may also be accelerated by
shocks associated with star formation, the large scale Galactic wind, or
activity at the centre of the Galaxy.

Diffusive shock acceleration (DSA) efficiently produces a $T^{-2}$ CR energy spectrum where $T$ is the CR energy in eV.
The predicted maximum CR energy produced by SNR shocks is close to a PeV, 
although it appears that
the historical supernova remnants (SNR) are unable to reach this energy since their shocks are already significantly
decelerated (Zirakashvili \& Ptuskin 2008, Bell et al 2013).
In order to explain the Galactic CR population it is essential not only
that CR protons should be accelerated to a few PeV 
but also that the CR should be able to escape the SNR
without large energy loss.
Bell et al (2013) showed that the highest energy CR escape upstream from the shock into the interstellar medium.
However most of the shock-accelerated CR, by energy content as well as number, 
are carried downstream
into the interior of the SNR.
In this paper we examine the fate of these lower energy CR as they are advected
into the SNR where they remain until the SNR slows and disperses
into the interstellar medium (ISM).
Once carried into the SNR interior CR lose energy adiabatically as the SNR
expands.
An individual CR accelerated early in the Sedov phase has a much reduced
energy by the time it is released into the Galaxy.
This is often perceived as a difficulty in explaining CR origins.
However, the CR energy lost by adiabatic expansion is in fact
re-used to drive the blast wave and accelerate a new generation of CR
at a later time.
Chevalier (1983a) derived a self-similar Sedov blast-wave solution that includes 
CR pressure.
He showed that the CR pressure dominates the thermal plasma pressure at the centre of the remnant
even if only a relatively small fraction of the available energy is given
to CR by the shock.
Because CR have a smaller ratio of specific heats ($\gamma =4/3$)
than thermal particles ($\gamma =5/3$), CR lose less energy during
adiabatic expansion.
Thermal particles preferentially lose energy as they drive the blast wave
and accelerate more CR,
whereas CR preferentially keep their energy for release into 
the ISM at the end of the SNR's life.

In this paper we extend Chevalier's self-similar model
to derive the CR energy spectrum 
and the maximum CR energy inside a blast wave.
We show that CR produced by SNR can be divided into two populations:
(A) CR with energies above $\sim 200$GeV that escape ahead of the shock during SNR expansion
(B) CR advected into the interior of the SNR during expansion and then 
released from the SNR at the end of its life to provide the Galactic CR component below $\sim 200$GeV.
Instead of limiting the efficiency of Galactic CR production,
adiabatic losses during SNR expansion increase the efficiency by filtering energy
from the thermal plasma into CR.
The underlying principles of the calculation
apply to any blast wave, possibly including any launched 
from the centre of the Galaxy or from
star forming regions.

Using the formulation developed by Bell et al (2013)
we derive energy spectra and energy densities of CR within the SNR and 
the total energy of CR 
released into the surrounding medium.
Bell et al (2013) showed that the maximum CR energy is determined by the growth rate of the instability amplifying 
the magnetic field needed to confine CR in the shock environment during acceleration.
The results derived using Bell et al (2013) differ from those derived on an assumption
that the energy density of the amplified magnetic field
is proportional to the kinetic energy density $\rho _0 u_s^2$ of 
plasma with density $\rho _0$ overtaken by a shock with velocity $u_s$
(eg Berezhko \& V\"{o}lk 2004, 2007).
CR produced by SNR can be divided into populations A and B as defined above.
The overlap of the two populations at an energy of about 200GeV may be related to the
break in the CR energy spectrum measured by PAMELA (Adriani et al, 2011)
and other experiments (Ahn 2010, Tomassetti 2012).

Throughout this paper we consider only proton acceleration.
Wherever CR are mentioned we refer to protons
unless otherwise stated.

\begin{figure*}
\includegraphics[angle=0,width=12cm]{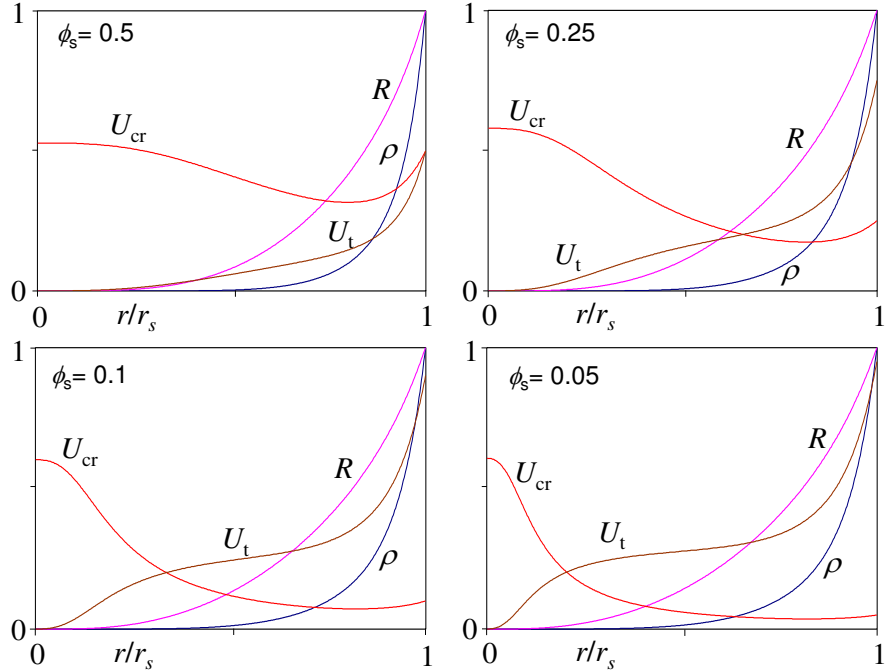}
\caption{
Profiles for different CR fractions $\phi _s$.
The axes are linear, not logarithmic.
}
\label{fig:figure1}
\end{figure*}


\section{Sedov self-similarity}

In this section we derive the Sedov self-similar solution including the CR
pressure as well as the thermal pressure.
Chevalier (1983a) has previously derived this self-similar solution
but we present the derivation in a form that facilitates calculation of
the self-similar CR energy distribution 
inside the blast wave.
A detailed time-dependent numerical study of the effect of efficient CR acceleration
on SNR dynamics in the Sedov phase can be found in Castro et al (2011).

The essential feature of the Sedov solution for an expanding blast wave
is that the total energy is conserved.
At any time during self-similar expansion 
into a uniform medium with density $\rho _0$
the energy in the blast wave is proportional
to $\rho _0 r_s^3 u_s^2$ since 
the energy density at any point inside the blast wave is proportional to $\rho _0 u_s^2$
(assuming that the shock Mach number is high)
and the volume is proportional to $r_s^3$
where $r_s$ is the radius of outer shock.
From energy conservation $r_s^3 u_s^2$ is constant,
so $r_s \propto t^{2/5}$ and $u_s \propto t^{-3/5}$.

In reality, and as part of this model, some energy is lost from the blast wave due to CR escaping upstream
as estimated below in equation 19.
{\bf 
If the energy loss is self-similar in the sense that the total blast wave energy $E$ decreases
in proportion to  $t^{-\beta}$ then $r_s \propto t^{(2-\beta)/5}$, $u_s \propto t^{-(3+\beta)/5}$,
and $\beta=-2(dE/dr_s)/(5(E/r_s)-(dE/dr_s))$.
If the energy loss due to CR escaping upstream is $0.03 \rho _0 u_s^3$ per unit shock area (Bell et al 2013)
then $ dE/dr_s \approx -0.13E/r_s$ for $E\approx 3\rho_0 u_s^2 r_s^3$, which gives $\beta\approx 0.05$, 
$r_s \propto t^{0.39}$ and $u_s \propto t^{-0.61}$ 
instead of $r_s \propto t^{0.4}$ and $u_s \propto t^{-0.6}$.
}
This will produce a very slight flattening in the CR spectrum since it reduces the energy given to low energy CR
later in the life of the SNR.  
Because the effect is small we neglect the effect of energy loss to CR and proceed on the assumption that
 $r_s \propto t^{0.4}$ and $u_s \propto t^{-0.6}$.

Self-similarity is independent of the ratio of specific heats $\gamma$.
It also holds for a mixture of CR and thermal gases with different $\gamma$
provided the CR acceleration efficiency is constant in time. 
We consider the case in which
the immediately post-shock CR pressure $P_{CR}$ is a fraction $\epsilon$
of the total post-shock pressure $P_s$ with the thermal
pressure $P_t$ providing the balance of the post-shock pressure:
$$
P_{CR}(r_s)=\epsilon P_s
\hskip 1 cm
P_t(r_s)=(1-\epsilon) P_s
\eqno{(1)}
$$
In reality, $\epsilon$ probably varies as the shock speed changes,
but for simplicity, and because it is unclear whether $\epsilon$ increases or decreases,
 we assume that it remains constant throughout the 
Sedov phase.
For convenience and usefulness in later sections of this paper 
we introduce $R(r)$ as the
radius of the shock when the fluid element presently at position $r$
was overtaken by the shock.
Since the mass presently inside the radius $r$ is equal to the mass inside
the shock when the shock was at radius $R$,
$$
\int _0 ^r 4 \pi \rho (r') r'^2 dr'
= \frac{4\pi}{3} \rho _0 R^3
\eqno{(2)}
$$
where $\rho$ is the present density profile.
$R(r)$ is a function of the present radius $r$.
Since $u_s \propto r_s^{-3/2}$
the post-shock pressure was $(R/r_s)^{-3}P_s $ when the fluid element
now at radius $r$ passed through the shock.
Hence the CR pressure at radius $r$ is reduced by adiabatic expansion to
$\epsilon P_s (R/r_s)^{-3} (\rho /\rho _s)^{4/3}$ where $\rho_s$ is the 
post-shock density.
Similarly the thermal pressure is 
$(1-\epsilon) P_s (R/r_s)^{-3} (\rho /\rho _s)^{5/3}$
so the total pressure at radius $r$ is
$$
P= P_s \frac{r_s^3}{R^3} \left [ 
(1-\epsilon) \left (\frac{\rho}{\rho _s} \right )^{5/3}
+\epsilon \left (\frac{\rho}{\rho _s} \right )^{4/3}
\right ]
\eqno{(3)}
$$
where all quantities are defined at the present time.
This equation assumes that all CR remain relativistic even as they cool adiabatically.
In practice mildly relativistic CR become non-relativistic as they cool adiabatically 
and their $\gamma$ changes from $4/3$ to $5/3$.
We neglect this effect under the assumption
that most of the CR energy resides in CR that
remain relativistic.
For example, if a  $T^{-2}$ spectrum extends to 1PeV, CR 
with a Lorentz factor less than two account for only 6\% of the total CR energy
and CR with a Lorentz factor less than ten account for 15\% of the total.
See Chevalier (1983a,b) for a discussion of this issue when the shock-accelerated spectrum is steeper than  $T^{-2}$.

We also assume that CR diffusion can be neglected and that
CR remain localised to the same fluid element after 
passing through the shock.
This is a good assumption for most CR,
since CR are spatially localised by their small Larmor radius:
the Larmor radius 
of a CR with energy $T_{GeV}$
in GeV in a 10$\mu $G magnetic field is only
$10^{-7}T_{GeV}$parsec.
Furthermore it is part of the theory of diffusive shock acceleration
that all except the very highest energy CR exit the
acceleration process by being advected away downstream with
the thermal plasma.
Hence advection dominates diffusion over most of the CR energy range, 
and diffusion can be neglected for bulk properties 
of CR such as the integrated energy density of all CR from the lowest to the
highest energy.

From self-similarity, $\rho = \rho (r/r_s)$,
the fluid velocity inside the SNR takes the form
$u=t^{-3/5} f(r/r_s)$,
and pressure takes the form 
$P=t^{-6/5} g(r/r_s)$
where $f$ and $g$ represent the shape of the velocity and
pressure profiles.
The self-similar equation for mass conservation is then
$$
\frac{u_s r}{r_s}
\frac {\partial \rho}{\partial r}
=\frac {1}{r^2} \frac {\partial (r^2 \rho u)
}{\partial r}
\eqno{(4)}
$$
where the left hand side of the equation is the self-similar
equivalent of $-\partial \rho /\partial t$.
The self-similar form of the momentum equation is
$$
\frac{u_s r}{r_s}
\frac {\partial u}{\partial r}
+\frac{3u_su}{2r_s}
=
u \frac {\partial u}{\partial r}
+\frac{1}{\rho}\frac {\partial P}{\partial r}
\eqno{(5)}
$$
where we have used equation 4 and the self-similar relation $u_s/r_s=2/5t$.
Equations 2-5 define the Sedov self-similarity solution
with the CR pressure included
where equations 2 and 3 combined represent energy conservation.  
The equation for mass conservation can be integrated to give
$$
\frac{u}{u_s}=\frac { \rho r^3 - \rho _0 R^3}{\rho r^2 r_s}
\eqno{(6)}
$$
The asymptotic solution close to the centre of the blast wave is
derived in appendix A:
$$
\frac {\rho}{\rho _s }
=
\left \{ 
\frac{1}{5(1-\epsilon)}
\left [ \left (16 \epsilon^2 + \frac{10 (1-\epsilon) \rho _s P_c r^3}{\rho _0 P_s r_s^3} \right )^{1/2}
- 4\epsilon \right]
\right \}^3
\eqno{(7)}
$$
where $P=P_c$ and $\partial P/\partial r=0$ at zero radius. 
For $\epsilon \rightarrow 0$ (negligible CR pressure)
$\rho \propto r^{9/2} $.
For non-zero $\epsilon $ (CR dominant at the centre) 
$\rho \propto r^9$ as $r \rightarrow 0$.
The asymptotic forms of $R$ and $u$ can be
derived from equations 3, 6 and 7.
Boundary conditions are imposed at the shock
where
$$
\rho_s = (4+3\epsilon ) \rho _0
\hskip 0.2 cm ; \hskip 0.2 cm
P_s= \frac {3+3\epsilon}{4+3 \epsilon} \rho _0 u_s^2
\hskip 0.2 cm ; \hskip 0.2 cm
R=r_s
\eqno{(8)}
$$
We solve the equations numerically 
by integrating towards the centre
from the shock radius $r_s$ 
until numerical accuracy is lost close to $r=0$ due to the density becoming very small
($\rho \propto r^9$ for small $r$). 
The profiles close to $r=0$ are derived from the asymptotic solution
given in equation 7 and fitted to the numerical solution by 
suitable choice of $P_c$.
The resulting profiles are given in figure 1 for various 
shock acceleration efficiencies (see also Tables 1 to 5 of Chevalier (1983a)).
We define $\phi $ as the ratio of the CR energy density
$U_{CR}$ to sum of the thermal $U_t$ and CR energy densities:
$\phi = U_{CR}/(U_t+U_{CR})$.
The subscript $s$ denotes the value at the shock.
$\epsilon$ and $\phi _s$ are related by
$$
\phi _s=\frac {2\epsilon}{1+\epsilon}
\hskip 0.2 cm ; \hskip 0.2 cm
\epsilon= \frac {\phi _s}{2-\phi _s}
\eqno{(9)}
$$
and the post-shock thermal and CR energy densities are
$$
U_{t,s}= \frac{18 (1-\phi_s )\rho_0 u_s^2}{(8- \phi _s)(2-\phi _s)}
\hskip 0.2 cm ; \hskip 0.2 cm
U_{CR,s}= \frac{18 \phi _s \rho_0 u_s^2}{(8- \phi _s)(2-\phi _s)}
\eqno{(10)}
$$

In figure 1 we see that the thermal energy density 
always decreases towards the centre of the 
blast wave.
In contrast, for all cases plotted in figure 1 the CR energy density
is greater at the centre than immediately downstream of the shock.
The central part of the blast wave can be characterised 
as a CR bubble with 
low thermal energy density and low mass density.
The radius of the CR bubble decreases as the CR fraction $\phi _s$
decreases,
but even when only 5 percent of the post-shock energy density is given to CR ($\phi _s=0.05$)
the CR bubble extends out to 10-20 percent of the shock radius.
Adiabatic expansion inside the blast wave acts as a filter which
transfers thermal energy to CR energy.

\begin{figure}
\includegraphics[angle=0,width=8cm]{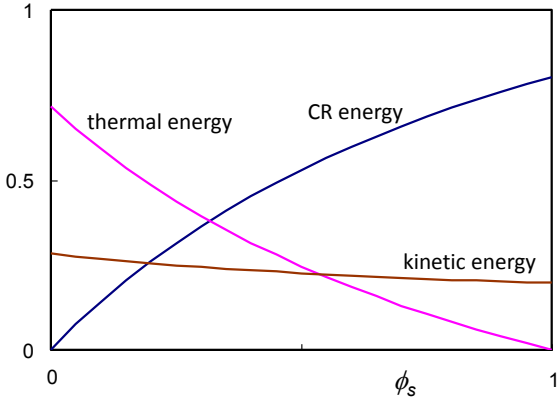}
\caption{
Fraction of total blast wave energy in CR, thermal and kinetic energy
as a function of the fraction of energy $\phi_s = U_{CR,s}/(U_{t,s}+U_{CR,t})$ 
given to CR at the shock.
See also Table 6 of Chevalier (1983a).
}
\label{fig:figure1}
\end{figure}

Figure 2 plots the CR, thermal and kinetic total energies as a function of $\phi _s$
(see also Table 6 of Chevalier (1983a)).
It shows that as much as 70-80\% of the total energy in the blast wave can be given to CR if
CR acceleration at the shock is highly efficient.
If equal energies are given to CR and thermal particles at the shock,
CR contribute 
53\% of the total blast wave energy (thermal+CR+kinetic).
Even if only 10\% of the total CR plus thermal energy at the shock is given to CR,
the CR energy in the blast wave is still 14\% of the total.
In the limit of small $\phi _s$, the total energy of CR
is $E_{CR}=1.5 \phi _s E_0 =3 \epsilon E_0$
where $E_0$ is the total energy of the blast wave.
Far from reducing the efficiency of CR production, the
hydrodynamics of the blast wave gives a proportion of the total blast wave energy
to CR which is greater than the fraction $\phi _s$ of energy given to CR at the shock.
Instead of being a problem for CR production, adiabatic expansion works to increase
the fraction of the supernova energy given to CR.
When a SNR finally disperses the CR energy released into the ISM
may be a large fraction of the energy of the original explosion.


\section{Approach to self-similarity}

$R(r)$ is the radius of the shock front at the time when the fluid element now at position $r$
was overtaken by the shock.
As seen in figure 1, a fluid element 
presently located about half way between 
the centre of the blast wave and the present shock radius
was overtaken by the shock when it was only $\sim 10 \%$
of its present radius.
Consequently the early non-Sedov evolution of the blast wave affects a large part of its
interior.
Self-similarity cannot be naively assumed even when the blast wave has expanded to $10\times $
or even $100\times $
the radius $r_{f}$ at which it completed the 
ejecta-dominated phase 
(sometimes known as the free expansion phase)
and entered the Sedov phase.
We characterise $r_f$ as the radius at which the swept-up mass 
$4 \pi \rho _0 r_f^3 /3$
is equal to the
ejected mass $M_{ej}$.
Figure 3 provides insight into the late-time effect of the early pre-Sedov
history of the blast wave.
It plots $R(r)$ and 
the shock velocity $u_s (r)$ defined as the velocity of the shock at the time
when the fluid element now at $r$ was overtaken by the shock.
Curves are plotted for different CR acceleration efficiencies:
$\phi _s =0.1$ and $\phi _s =0.5$.
$u_s$ is plotted relative to its present value at $r=r_s$.
For both values of $\phi _s$
the figure shows that a fluid element now at radius $r_s/2$
passed through the shock when the shock velocity was $\sim 30\times$
its present value.
For example if the present shock velocity is $100{\rm km\ s}^{-1}$ a fluid element
at radius $r_s/2$ would have passed through the shock when its velocity was 
$\sim 3000{\rm km\ s}^{-1}$.
Fluid elements close to the centre of the blast wave would have been shocked
at unrealistically high velocities.
This casts doubt on the realism of the Sedov solution for the inner parts
of the blast wave.

\begin{figure}
\includegraphics[angle=0,width=8cm]{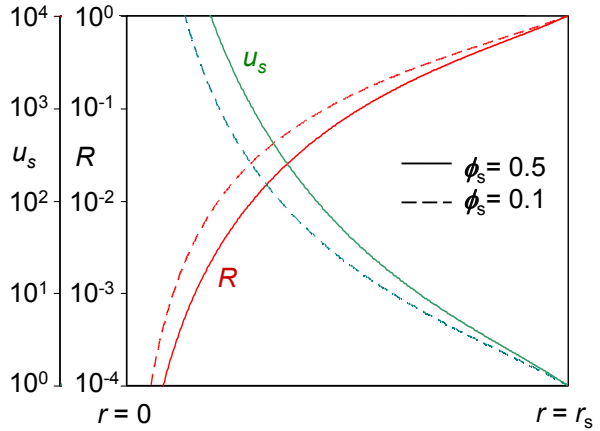}
\caption{
The curve for $R$ plots the radius, relative to the present shock radius, at which
fluid elements original passed through the shock.
Similarly the curve for $u_s$ plots the shock velocity,
relative to the present shock velocity,
when the fluid element now at radius $r$ passed through the shock.
}
\label{fig:figure3}
\end{figure}

We examine the approach to Sedov self-similarity by time-dependent
Lagrangian hydrodynamic calculation of a blast wave driven by a thin spherical shell
with mass $M_{ej}$
initially expanding into a uniform medium of density $\rho _0 $ with velocity $u_{ej}$.
 In reality the hydrodynamic structure of the early ejecta-dominated phase is much more complicated
(Chevalier 1982, Truelove \& McKee 1999).  Ejecta are launched with a range of velocities
 rather than a single velocity $u_{ej}$ but the thin shell model provides guidance on the
 validity of the Sedov model that is our concern here.
The solution converges to the self-similar Sedov solution when the shock radius $r_s$ is much greater
than the radius $r_f$.
The comparison is shown in figure 4 for $\phi _ s=0.25$ where the profiles of the
mass density $\rho$, the CR energy density $U_{cr}$ and the thermal energy density
$U_t$ are plotted when the blast wave has expanded to 10, 100 and 1000 times the
radius $r_f$.
The density profile is nearly unaffected by the pre-Sedov history.
The energy densities are nearly unaffected when the 
blast wave has expanded by a factor of 1000 in radius,
but strongly affected when the blast wave
has expanded by a factor of 10.
However, for all values of $r_s/r_f$ in figure 4,
the CR energy density exceeds the thermal energy density
in the inner half (by radius) of the blast wave.
Hence the conclusion of section 2 still stands that
adiabatic expansion acts to filter energy into CR
and the inner parts of the blast wave are dominated by
CR pressure.
The total pressure (CR plus thermal) at the centre of the blast
wave is approximately independent of $r_s/r_f$ in figure 4 since it is determined by
the need to drive the blast wave into the surrounding medium.

In passing we note that the agreement between the curve for
$r_s/r_f= \infty$ and $r_s/r_f=1000$ in figure 4 for all except small radius
where they would be expected to differ is evidence that both the self-similar and
the thin shell calculations are reliable since the
the curves were calculated with different computer codes
using different numerical methods.

\begin{figure}
\includegraphics[angle=0,width=8cm]{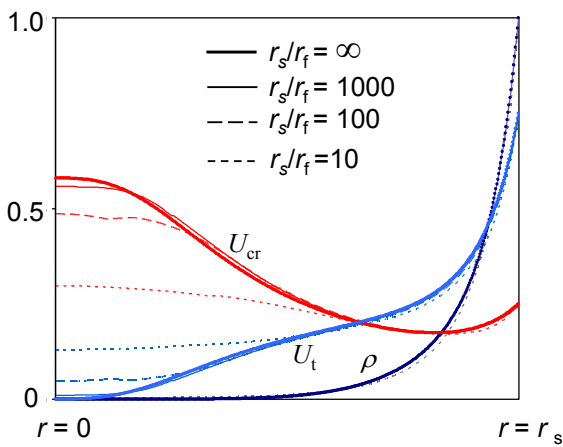}
\caption{
Profiles of the mass  density, CR energy density and thermal energy density for $\phi _s=0.25$.
The self-similar solution is represented by the thick continuous lines.
The other lines represent the approach towards self-similarity as the SNR radius $r_s$
increases and an initial ejecta-dominated phase passes into history. 
}
\label{fig:figure4}
\end{figure}


\section{The maximum CR energy inside the blast wave}

In this section we derive the maximum CR energy as a function of radius.
We will assume self-similarity in this section and then
examine effects arising from the pre-Sedov history
in section 5.
A fluid element presently at radius $r$ passed through the shock
when its radius was $R$.
We assume that the CR accelerated by the shock followed a $T^{-2}$ energy
spectrum up to a maximum CR energy $T_s (R) $ in eV.
After adiabatic expansion, 
the CR spectrum is still proportional to $T^{-2}$,
but the maximum CR energy at radius $r$
is reduced to
$$
T_{max}(r)= T_s(R) \left ( \frac{\rho}{\rho _s} \right )^{1/3}
\eqno{(11)}
$$
$T_s(R)$ is determined by the microphysics of CR acceleration 
and the CR-driven amplification of magnetic field in the shock precursor.
We consider three different models (A, B \& C) for $T_s(R)$
as follows.

Equation 7 depends on the assumption that CR diffusion is small.
While diffusion has negligible effect on bulk CR properties such
as the CR energy density, as discussed in section 2,
it could be more important for CR with energy $T_{max}$ which have a relatively large 
Larmor radius.
As discussed below the maximum CR energy in the centre of a SNR at the
end of its life is about 10TeV and these have a Larmor radius in a
10$\mu$G magnetic field of 0.001parsec which is very much less
than the SNR radius during the Sedov phase.
Consequently, CR diffusion inside old SNR can only be important if the interior
magnetic field is very small, and even then
CR would be unable to escape through the
larger compressed interstellar magnetic field closer to the shock.
At early times during the Sedov phase $T_{max}$ is larger but the magnetic field
is also larger due to field amplification.
Neglect of diffusion therefore seems reasonable,
but the validity of the assumption might be tested
with more complete calculations.

The dependence of $T_{max}(r)$ on $\rho (r)$ as given in equation 11 determines the maximum
CR energy inside the blast wave for a given maximum CR energy $T_s(R)$ at the shock.
Bell et al (2013) showed that $T_s(R)$ in the early evolution of an SNR 
is determined by the growth rate of the instability that amplifies the magnetic field.
Model C below is based on this understanding, but firstly for comparison
we consider two other models, A and B, based on simpler ways of estimating $T_s(R)$
at the shock.
Model A neglects magnetic field amplification during acceleration and assumes Bohm diffusion.
Magnetic field amplification is well attested by observation as well as theory
so Model B includes magnetic field amplification but still assumes Bohm diffusion. 
Model C both includes magnetic field amplification and avoids assuming Bohm diffusion.

{\bf Model A:} Firstly we consider the option that $T_s=u_s B_0 r_s/8$ which is derived from Lagage \& Cesarsky
(1983a,b) where $B_0$ is the upstream magnetic field 
(ie no magnetic field amplification ahead of the shock).  
This expression for $T_s$ is based on Bohm diffusion 
(defined here as $D_{Bohm}=r_gc$ where $r_g$ is the CR Larmor radius) in a magnetic field $B_0$ during shock acceleration.
The factor $1/8$ assumes that CR spend equal times upstream and downstream
during acceleration (Bell 2013).
Apart from the factor $1/8$ this is also the maximum CR energy
derived by Hillas (1984) for generalised CR acceleration.
For self-similar expansion $u_s \propto r_s^{-3/2}$.
Here and throughout the rest of the paper, for
a SNR approaching the end of its life,
we assume the following standard values:
$$B_0=5\mu G 
\hskip 0.5 cm
u_s=30 {\rm km\ s}^{-1}
\hskip 0.5 cm
r_s=100{\rm pc}.
$$
The maximum CR energy at a radius $r$ inside the blast wave is then
$$
T_{max}(r)= 5 \left (\frac {R}{r_s}\right )^{-1/2} \left ( \frac{\rho}{\rho _s} \right )^{1/3}\ {\rm TeV}
\eqno{(12)}
$$
as plotted in figure 5, where $R$ and $\rho$ are functions of $r$.
The curves labelled `model A' in figure 5 show that 
the maximum CR energy falls away slowly inside the blast wave,
but remains greater than 1TeV until very close to the centre.
With this recipe for the magnetic field, CR released into the ISM when a SNR 
reaches the end of its life can only replenish the Galactic CR population 
up to energies of a few TeV.

{\bf Model B:} As pointed out by Lagage \& Cesarsky (1983a,b)
CR cannot be accelerated to PeV energies
if the magnetic field at the shock is limited to interstellar values of a few $\mu$G.
Magnetic field amplification (Bell 2004) facilitates CR acceleration to PeV energies.
Option C will apply the latest theories of magnetic field amplification, but before that
we consider the case in which $T_s=u_s B r_s/8$ and the pre-shock magnetic field is amplified such that 
the magnetic energy density at the shock is a fixed fraction of
the available energy, $B^2/2\mu _0= \xi \rho_0 u_s^2$.
V\"{o}lk et al (2005) suggest from observations that the downstream magnetic energy density is
$\sim 3$\% of  $\rho_0 u_s^2$ implying $\xi \sim 0.003$
(depending on the magnetic field orientation) when allowance is made for
magnetic field compression at the shock (increasing $B^2$ by $\sim 10$) 
when estimating the upstream magnetic field.
In this case,
the upstream magnetic field is the maximum of the amplified field and 
a typical ISM field of  $5 \mu {\rm G}$
$$
\frac{B}{\mu {\rm G}}=\max \left [ 5\ , \ 1.2 \left ( \frac{\xi }{0.003} \right ) ^{1/2}
\left ( \frac{n_e }{{\rm cm}^{-3}} \right ) ^{1/2}
\left ( \frac{u_s }{30 {\rm km \ s}^{-1}} \right ) 
\right ]
\eqno{(13)}
$$
The corresponding radial profile of the maximum CR energy for 
$B_0=5\mu G$, 
$u_s=30 {\rm km\ s}^{-1}$, and
$r_s=100$pc
is
$$
T_{max}(r)= \max
\left [
5.6 \left (\frac {R}{r_s}\right )^{-1/2} \ , \ 
1.4 \left (\frac {R}{r_s}\right )^{-2}
\right ]
\left ( \frac{\rho}{\rho _s} \right )^{1/3}\ {\rm TeV}
\eqno{(14)}
$$
as plotted in the curves labelled `model B' in figure 5.
$T_{max}$ falls away for a small distance inside the shock before
increasing dramatically at the centre of the SNR due to 
magnetic field amplification.

\begin{figure}
\includegraphics[angle=0,width=8cm]{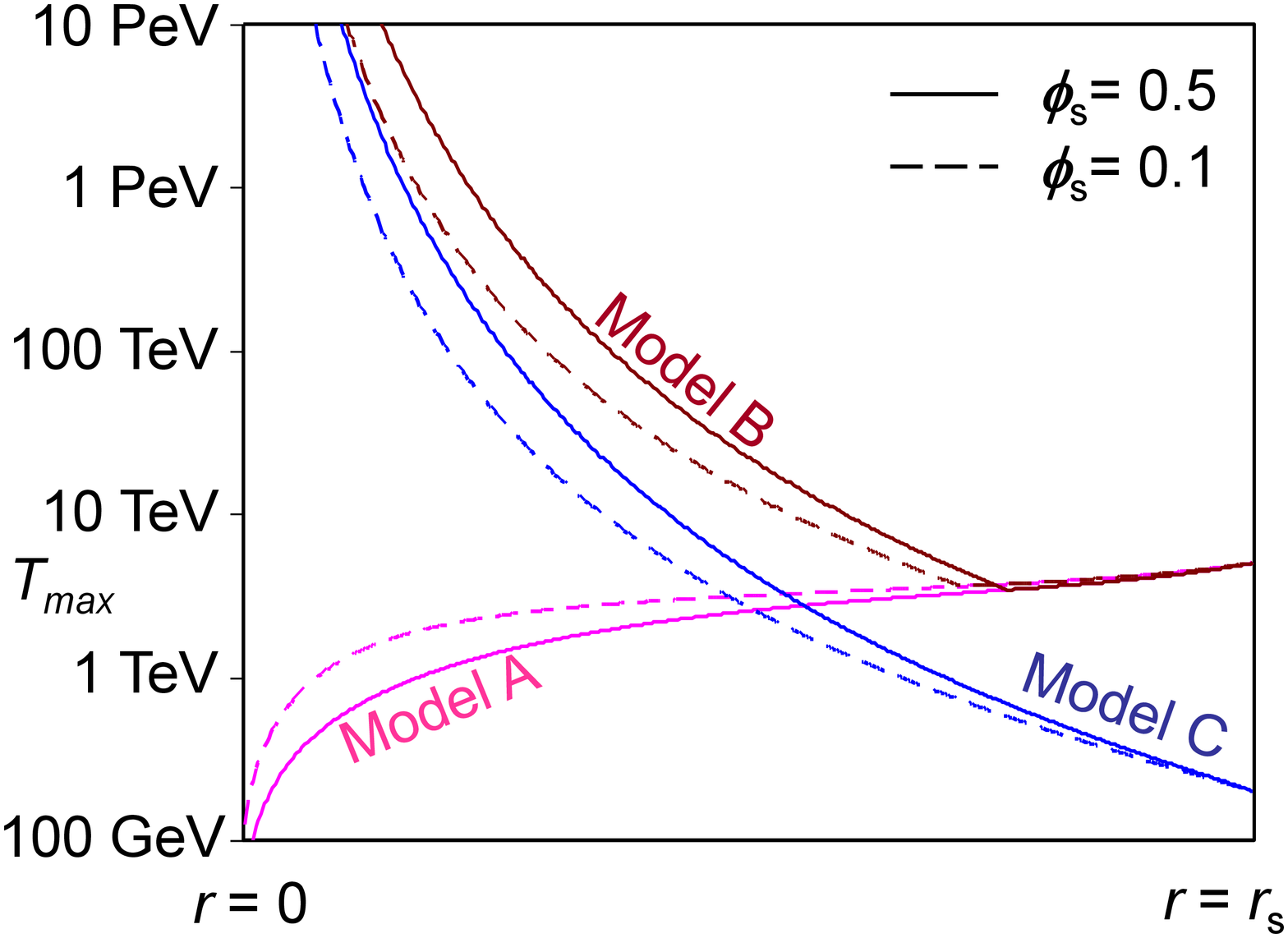}
\caption{
Maximum CR energy as a function of radius at the end of a SNR lifetime 
as given by Models A, B \& C for $\phi _s =0.5$ (full line)
and $\phi _s =0.1$ (dashed line).
}
\label{fig:figure5}
\end{figure}

{\bf Model C:} 
Model C is based on our current best understanding of
magnetic field amplification as presented in Bell et al (2013).
Models A and B based the calculation of $T_{max}$ on the assumption of Bohm diffusion
in a magnetic field that was unamplified in model A or deduced from observation in model B.
Bell et al (2013) showed that the maximum CR energy in young SNR
is more reliably determined by 
the growth rate of the instability responsible for magnetic field amplification.
The non-resonant hybrid (NRH) instability dominates in young SNR.
Its maximum growth rate is proportional to the electrical current
carried by CR escaping upstream of the shock:
$\gamma _{max}=0.5 j_{CR} \sqrt{\mu _0/ \rho}$.
For a given CR energy flux, the CR electric current is inversely proportional
to the CR energy.
This imposes a limit on energy to which CR can be accelerated since CR
with very high energy carry a very small electric current
for a given CR energy flux.
A requirement of 5 e-foldings at the maximum growth rate
means that the CR charge per unit area
$j_{CR} \tau$ escaping a SNR with age $\tau$ must exceed 
$10 \sqrt{\rho/\mu _0}$.
By this argument, Bell et al (2013) derived an estimate (their equation 21) for $T_{max}$:
$$
T_{max}=230 \eta_{0.03} n_e^{1/2} u_7^2 R_{pc}\ {\rm TeV}
\eqno{(15)}
$$
where $u_7$ is the shock velocity in units of 10,000 km s$^{-1}$,
$R_{pc}$ is the shock radius in parsec,
$n_e$ is the electron density in cm$^{-3}$
and $\eta _{0.03}$ is an efficiency factor normalised 
relative to $\eta=0.03$ as defined by Bell et al (2013)
such that
it is reasonable to assume that  $\eta _{0.03}=1$. 
This expression for $T_{max}$ was based on the assumption that the
magnetic field is strongly amplified by the NRH instability.
This is correct for shock velocities greater than about 1,000 km s$^{-1}$ 
where the ${\bf j}_{CR} \times {\bf B}$ forces exerted by the
CR current on the thermal plasma exceed the magnetic force
$-{\bf B}\times (\nabla \times {\bf B})/\mu_0$ acting 
within the thermal plasma.
At shock velocities less than $\sim$1,000 km s$^{-1}$ 
the NRH instability is inactive (Schure \& Bell 2013) and magnetic fluctuations are
excited by the resonant Alfven instability (Lerche 1967, Kulsrud \& Pearce 1969, Wentzel 1974)
that generates Alfven waves with a wavelength $2\pi /k$ matching the CR Larmor radius $r_g$.
The Alfven instability operates differently from the NRH instability and dominates
in a different regime but its maximum growth rate is a numerical factor times
$0.5 j_{CR} \sqrt{\mu _0/ \rho}$.
The numerical factor is close to one 
as noted by Zirakashvili \& Ptuskin (2008)
but depends upon
the form of the CR energy distribution
(see Appendix B).
Hence the argument based on the NRH instability (Bell et al 2013) for high
velocity shocks also applies to the Alfven instability at low velocity shocks,
and equation 15 can be applied to SNR
throughout the Sedov phase.
The corresponding profiles of $T_{max}$ inside the blast wave are plotted
as the curves labelled `model C' in figure 5.
$T_{max}$ at a radius $r$ is calculated from equation 15 with 
 $R_{pc}$  and $u_7$ set to the shock radius and shock velocity when the fluid element
 at $r$ was overtaken by the shock.
 
 The results obtained with models A, B \& C are discussed further in the next two sections.

\section{T$_{max}$ near the centre}

According to figure 5 the maximum CR energy $T_{max}$ 
is unbounded at zero radius in models B and C.
This is an artifact due to the projection of self-similar Sedov expansion
back to zero SNR radius.
In the pre-Sedov ejecta-dominated phase, the shock velocity is much lower
than that given by the Sedov model in which the expansion velocity 
is infinite at $t=0$.

In section 3 (see figure 4) the effects of initial ejecta-dominated were
estimated using a time dependent model in which the shock was driven
by a thin shell representing the ejected mass $M_{ej}$.
The same thin-shell model can be used to estimate $T_{max}$ near the centre of the blast wave
where the history of the ejecta-dominated phase is important.
The shock velocity is nearly constant during the ejecta-dominated pre-Sedov phase 
and the radius is small initially so $T_{max}$ 
turns over on approaching the centre of the blast wave 
as plotted in figure 6 in accord with equation 15.
Nevertheless, $T_{max}$ at the centre of the blast wave 
can be $\sim 100-1000$ times larger than $T_{max}$ at the shock.
In old SNR ($r_s/r_f =100-1000$) CR energies may
reach $\sim 10-100$TeV in the centre of an SNR even though
CR are currently accelerated only to $\sim 100$GeV at the shock.
Early in the Sedov phase ($r_s/r_f \sim10$), $T_{max}$ at the centre of the blast
wave is only $\sim 10$ times larger than $T_{max}$ at the shock.

\begin{figure}
\includegraphics[angle=0,width=8cm]{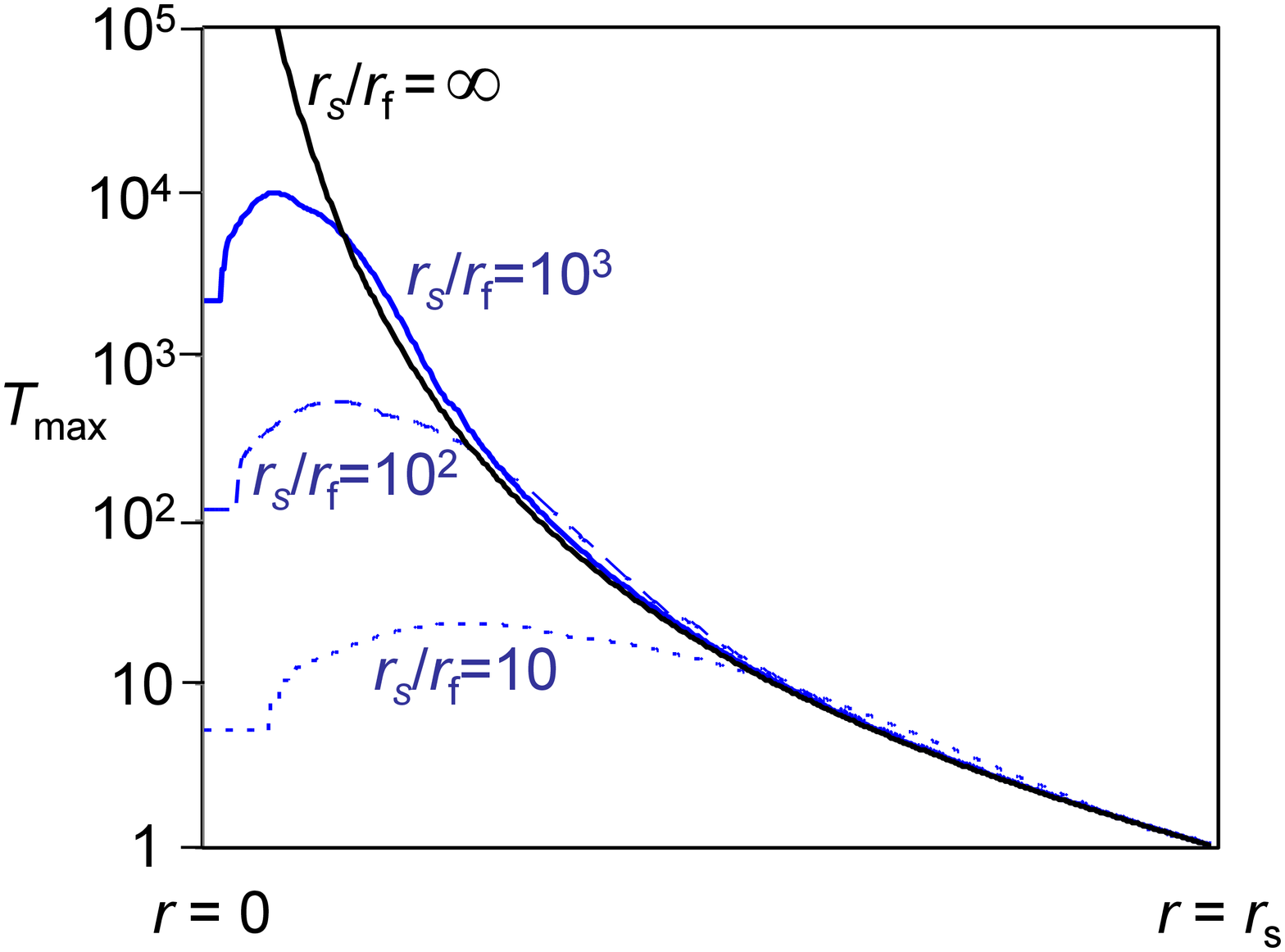}
\caption{
$T_{max}$ at radius $r$ within the SNR 
relative to $T_{max}$ at the shock
for model C with $\phi_s=0.25$.
$r_s/r_f$ is the ratio of the SNR radius $r_s$ to the radius $r_f$
at the end of the ejecta-dominated phase.
The Sedov self-similar result is given by 
$r_s/r_f= \infty $.
} 
\label{fig:figure6}
\end{figure}

\section{The limitations of models A and B} 

Models A and B predict larger CR energies than model C in the outer
parts of the blast wave
because they incorrectly assume Bohm diffusion in old SNR
when the Alfven instability is weakly driven.
Bohm diffusion occurs when CR trajectories are scattered with a mean
free path equal to the CR Larmor radius.
This is only possible if rapidly growing plasma instabilities 
produce large fluctuations in the field on the scale of a Larmor radius.
If the magnetic field remains essentially uniform on the Larmor scale
the CR are unscattered and diffusive shock acceleration is too slow 
for CR to reach the energy $u_sB_0 r_s/8$ assumed in models A and B.
Model C takes account of the instability growth time 
and consequently predicts the lower CR energies seen at large radius
in figure 5.
Models A and B therefore overestimate the CR energy at $r \sim r_s$
at the end of a SNR lifetime.

Model B also overestimates the maximum CR energy
in the centre of the blast wave.
Model B assumes that Bohm diffusion applies and that
the Bohm diffusion coefficient should be calculated from the total magnetic field.
In reality, fluctuations in the magnetic field grow on a wide
range of scales from the Larmor radius of GeV protons
to the Larmor radius of the highest energy CR.
Bohm  diffusion depends on a match between the Larmor radius of
the scattered CR with the scalelength of the magnetic field.
Only components of the magnetic field structured on the scale of the CR 
Larmor radius are effective in scattering a particular CR.
The magnetic field derived from x-ray synchrotron emission at the shock
(Berezhko et al 2003, Vink \& Laming 2003, V\"{o}lk et al 2005) 
is the total magnetic field.
The component of the magnetic field on the Larmor radius of
a particular CR is smaller.
Model B uses the observed magnetic field 
as calculated by V\"{o}lk et al (2005) to predict $T_{max}$ 
and therefore model B overestimates the maximum CR energy.
Compensation for this effect would probably reduce $T_{max}$
in agreement with model C.


\section{The CR energy spectrum}

The maximum CR energy $T_{max}$ is plotted in 
figures 5 and 6 for different models as a function of radius $r$.
Working on the basis that the energy spectrum at any radius 
follows a $T^{-2}$ power up
to the local maximum CR energy $T_{max}(r)$ 
we integrate in radius to calculate 
the differential energy spectrum of
the total CR population inside the blast wave.
The full lines in figure 7 present the CR spectrum calculated for
model C for two different CR acceleration efficiencies, 
$\phi_s$ equal to 0.1 and 0.5.
The CR energy $T$ is normalised to $T_{max,s}$ which is the 
current value of $T_{max}$ at the shock.
The spectrum is proportional to $T^{-2}$ for $T<T_{max,s}$ since this 
power law applies at all points inside the blast wave in this energy range.
The local maximum CR energy $T_{max}$ increases towards the centre of the remnant
so CR reach the highest energies only in a small volume close to the centre.
Consequently the spectrum is steeper for $T>T_{max,s}$ but still follows a power law. 
The spectral index of 2.6 for $T>T_{max,s}$ is close to that of Galactic CR up to the knee, but
this must be coincidental since the spectrum of CR arriving at the Earth
is expected to be steepened by energy-dependent losses during propagation from the source.
The shape of the spectrum is nearly independent of $\phi _s$, but slightly flatter
for $\phi _s=0.1$.

The self-similar spectrum calculated for model C 
extends without limit towards infinite CR energy,
representing CR acceleration by an infinitely fast Sedov blast wave 
expanding from a central singularity.
The dashed curves in figure 7 plot the spectrum calculated with the time dependent code
that models the pre-Sedov phase as described in sections 3 and 5.
This more realistic thin-shell model of early expansion causes the CR spectrum to terminate instead of extend to infinite energy.
The radius of a SNR expands by about 50 during the Sedov phase ($r_s/r_f\sim 50$)
in which case the spectrum terminates at $\sim 200 T_{max,s}$ at the end
of the Sedov phase.  In other words, towards the end of the Sedov phase,
CR near the centre of the SNR reach energies
which are about 200 times larger than the maximum CR energy at the shock.

The total CR spectrum inside the SNR has two important energies:
(i) the energy $T_{s,max}$ which is the maximum CR energy at the shock and at which
the spectral index steepens from 2.0 to 2.6,
(ii) $T_{t,max}$ which is the maximum CR energy anywhere in the SNR and
the energy at which the spectrum terminates.

From equation 15, $T_{max,s}=230 n_e^{1/2} u_{7}^2 R_{pc} {\rm TeV}$ where $R_{pc}$ and $u_7$
are the shock radius and the shock velocity.
The total energy of a Sedov blast wave is
$E= 3 \rho_0 u_s^2 R^3$ for $\gamma =5/3$ (the case of negligible CR pressure),
so this formula can be re-cast as 
$T_{max,s}=400 n_e^{1/6} E_{44}^{1/3} u_7^{4/3} {\rm TeV}$
where $E_{44}$ is the blast wave energy in units of $10^{44}{\rm J}$.
CR are only confined at the shock if the shock velocity is greater
than the Alfven speed $v_A$ since the resonant Alfven instability 
is only excited by CR drifting faster than the Alfven speed.
$v_A=10 B_5 n_e^{-1/2} {\rm km \ s}^{-1}$
where $B_5$ is the magnetic field in units of $5\mu {\rm G}$.
We make the assumption that CR are released into the ISM when the
Alfven Mach number decreases to 3 ($u_s=3v_A$)
in which case $u_7= 0.003B_5 n_e^{-1/2}$
and 
$$
T_{max,s}=200 n_e^{-1/2} E_{44}^{1/3} B_5^{4/3} {\rm GeV}
\eqno{(16)}
$$
Under these assumptions, and with our standard parameters,
CR are released into the ISM 
from the interior of the SNR at the end of its life with a power law spectrum
$T^{-2}$ at energies less than 200 GeV.
At energies above 200GeV the spectrum is steeper inside the blast wave and proportional
to $T^{-2.6}$.
The estimate of 200 GeV as the maximum energy to which CR are accelerated at the end of
the SNR lifetime will be reduced 
if collisional damping in a dense partially ionised plasma inhibits
the growth of CR-driven Alfven waves
as may be the case for the middle-aged SNR W44 where the proton spectrum
turns over at only 10GeV (Abdo et al 2010).

The energy $T_{max,t}$ at which the CR spectrum terminates
can be estimated as follows.
CR with the highest energy $T_{max,t}$ 
are found near the centre of the SNR (see figure 6).
In the thin shell model they were accelerated as the SNR entered the Sedov phase.
At that stage the shock velocity $u_f$ was around 
$10,000 {\rm km \ s}^{-1}$ ($u_{f7}=1$),
the maximum CR energy was 
$T_{max,f}=400 n_e^{1/6} E_{44}^{1/3} u_{f7}^{4/3} {\rm TeV}$
and the post-shock CR pressure was
$P_f \approx 0.75 \epsilon \rho _0 u_f^2$.
These initially very high energy CR cooled to the energy $T_{max,t}$ as the SNR expanded.
By the time the CR are released into the ISM their pressure
has decreased to $P_c\approx 0.25 \rho _0 u_s^2$ which is the pressure at the centre
of a Sedov blast wave expanding at velocity $u_s$.
Since individual CR energies reduce adiabatically in proportion to the CR pressure to
the power $1/4$, 
$
T_{max,t}= T_{max,f} (P_c/P_f)^{1/4}
= T_{max,f} (u_s/u_f)^{1/2} (3\epsilon )^{-1/4}
$.
For our standard values, 
$T_{max,f}\approx 400$TeV,
$u_f\approx 10,000 {\rm km \ s}^{-1}$,
$u_s\approx 30 {\rm km \ s}^{-1}$,
and $(3\epsilon )^{-1/4}\approx 1$,
giving
$T_{max,t} \sim 20$TeV.
The implication of this very approximate estimate is that
the maximum energy of CR released into the ISM from the
interior of an SNR is about 20TeV.
Their energy is much less than the $\sim 400$TeV 
to which they were originally accelerated in our simple model,
and their numbers are relatively small because of the
steeper energy spectrum ($T^{-2.6}$) above 200GeV.

\begin{figure}
\includegraphics[angle=0,width=9cm]{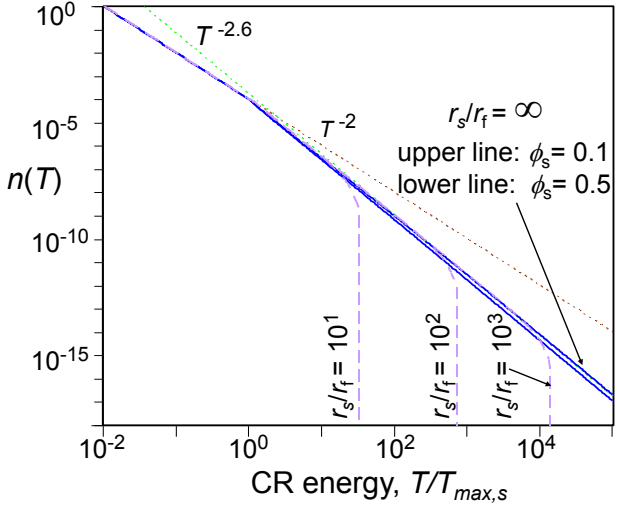}
\caption{
Energy spectra $n(T)$ integrated over all CR inside the blast wave as 
given by model C.
The full lines refer to the self-similar model with $r_s/r_f=\infty$
and with $\phi _s =0.5$
and $\phi _s =0.1$.  
The dashed lines 
refer to the thin shell model with $\phi_s=0.25$ and with the 
shock radius $r_s$ equal to 10, 100 and 1000 times the ejecta-dominated radius $r_f$.
The CR energy $T$ is normalised to the current $T_{max,s}$ at the shock.
The dotted lines are inserted as reference lines for $T^{-2}$ and $T^{-2.6}$ power laws.
}
\label{fig:figure7}
\end{figure}


\section{Galactic CR}

In this paper we
have shown that adiabatic losses do not reduce the total CR energy released into the ISM.
Any energy lost by CR due to adiabatic expansion is used to drive the blast wave and
accelerate a new generation of CR.
In fact, adiabatic processes increase the overall efficiency of CR production.
Losses due to adiabatic expansion are stronger for the thermal plasma ($\gamma=5/3$) 
than for CR ($\gamma=4/3$).
As shown in figures 1 \& 2, most of the energy in the blast wave can be given to CR.
The blast wave acts as a filter to accumulate CR which are then released into the ISM
as the SNR eventually dissipates.

Adiabatic expansion operates to increase the total SNR energy passed to CR
but it works against the production of CR with high energies reaching the knee in the spectrum.
As estimated in section 7, the maximum CR energy $T_{max,t}$ inside the SNR at the end of of its life is
of the order of 20TeV.
SNR in the late Sedov phase may efficiently produce the Galactic CR population up to the
maximum energy $T_{max,s}$ of CR being accelerated by the shock at the end of the SNR's life.
As shown in figure 7, the CR spectrum inside the SNR steepens at this point before terminating at
$T_{max,t}$.
It was shown by Bell et al (2013) and Schure \& Bell (2013) that CR above 200 GeV can instead be produced 
efficiently by young SNR,
but these are released into the Galaxy by escaping upstream
without passing into the interior of the SNR.
They are the highest energy CR being accelerated by the shock at any time by the expanding SNR.
They have long scattering mean free paths and carry the electrical current
needed to excite instabilities upstream of the shock.

CR accelerated by SNR can therefore be divided into two populations.
A high energy population (population A), extending from $\sim 200 {\rm GeV}$
to $\sim 1 {\rm PeV}$, escapes upstream with a $T^{-2}$ energy spectrum when averaged over the
Sedov phase.
A low energy population (population B), with a $T^{-2}$ energy spectrum below $\sim 200 {\rm GeV}$
and $T^{-2.6}$ between  $\sim 200 {\rm GeV}$ and  $\sim 20 {\rm TeV}$,
is released into the ISM by old SNR after residing inside the remnant between
acceleration and release.
Although both populations contribute Galactic CR between $\sim 200 {\rm GeV}$ and  $\sim 20 {\rm TeV}$,
population A increasingly dominates toward the higher end of this range because of its
flatter spectrum.

The production of the two CR populations is strongly related 
and they both have the same spectral index under the assumption that
shock acceleration produces a $T^{-2}$ spectrum.
However their history between acceleration and release into the
ISM is different so they may not connect seamlessly at the cross-over energy at $\sim 200 {\rm GeV}$.
We assess the connectivity of the two populations by comparing the energy released into the ISM
in each population.

Initially we compare the energy of each population in the limit of low acceleration efficiency
in which $\epsilon $ is small.
From section 2 and figure 2 the energy in low energy CR, population B, is 
$$ E_{B}\approx 3 \epsilon E_0
\eqno{(17)}
$$ 
for small $\epsilon$ where $E_0$ is the total blast wave energy.

The energy released into the ISM as population A can be estimated from equations
2-4 from Bell et al (2013) in which
CR escape ahead of the shock at energy $T_{max}$ with electric current $j_{CR}$ 
and consequent energy flux $j_{CR} T_{max}$.
From these equations, the rate of CR energy escape from unit surface area of the shock is
$0.75 P_{CR} u_s/ \log(eT_{max}/m_p c^2)$
where $P_{CR}$ is the CR pressure at the shock, $u_s$ is the shock velocity and
$eT_{max}/m_p c^2$ is the Lorentz factor of escaping CR protons.
The total energy $E_A$ released into the ISM with population A can be estimated by integrating
over CR released as the SNR expands from the radius $R_f$ at the beginning of the Sedov phase
to a radius $R_s$ when CR are released into the ISM, giving
$$
E_A=\int _{Rf}^{Rs} 4\pi R^2 \frac{ 0.75 P_{CR} u_s} { \log(eT_{max}/m_p c^2)}\ dt
\hskip 5 cm
$$
$$
\hskip 2 cm
\approx \frac{3 \pi}{4} \frac {\log (R_s/R_f) } {\log(eT_{max}/m_p c^2)}\ \epsilon E_0
\eqno{(18)}
$$
$eT_{max}/m_p c^2=10^6$ for acceleration to 1PeV at the beginning of the Sedov phase,
and $R_s=50R_f$ for deceleration from $10,000{\rm km \ s}^{-1}$ to $30{\rm km \ s}^{-1}$
during the Sedov phase in which $R_s \propto u_s^{-2/3}$, giving
$$
E_A \approx 0.7 \epsilon E_0
\eqno{(19)}
$$
where $E_0=3\rho_0 u_s^2 R_s^3$.
These estimates (equations 17 \& 19) gloss over a number of complicating factors, 
but they are sufficient to suggest
that the energies $E_A$ and $E_{B}$ in each population are comparable 
except that the energy in the higher energy population A
is probably $\sim 0.25$ times that in population B.
Hence the connection at around 200GeV can be expected to be reasonably smooth.
If we take our estimates of $E_A$ and $E_{B}$ at face value,
the spectrum at source below 200GeV is proportional to $T^{-2}$.
At energies a little above 200GeV the spectrum steepens to $T^{-2.6}$ as seen in figure 7
before flattening again to $T^{-2}$ as population A begins to dominate.
Of course, the spectrum of CR arriving at the Earth is steepened due to energy losses during propagation.
Also, the spectrum at source may deviate from a $T^{-2}$ spectrum 
as discussed for example by Bell et al (2011).

The formula $E_{B}=3 \epsilon E_0$ is correct for small $\epsilon$ and $\phi _s$.
If CR acceleration is more efficient and $\phi_s=0.5$ then the formula overestimates $E_{B}$
by a factor of 2 (see figure 2). 
$E_{B}$ and $E_A$ are then closer in value but still $E_{B}>E_A$
and the overall picture of the Galactic CR spectrum is more or less unchanged.

Adriani et al (2011) find evidence in PAMELA data for a flattening in the Galactic
CR spectral index above 200GeV.
The detailed spectral structure observed at 200GeV might be open to question
in the light of AMS data (Ting et al 2013),
but the change in index is supported by other data (Ahn et al 2010, Tomassetti 2012).
Our model suggests that the structure at 200GeV
might be due to the joining of population A and population B.
If anything we predict a local steepening above 200GeV
rather than a flattening of the spectrum,
and our prediction of the join occurring at 200GeV is uncertain easily by a factor of 2.
However, it appears very likely that Galactic CR at PeV  and GeV energies must 
have been accelerated at very different stages of SNR evolution
and escaped into the interstellar medium by different routes at different times.
More detailed modelling and observation is needed to establish whether the measured
structure in the Galactic CR spectrum can be explained by our model
or whether the answer lies in energy-dependent CR propagation from the SNR
to the Earth as proposed for example by Blasi et al (2012) or Tomassetti (2012)
or in spectral concavity due to non-linear effects as proposed by Ptuskin et al (2013).


\section{Observational consequences}

Finally we briefly note some observational consequences for SNR and other blast waves.
Our analysis predicts the existence of CR bubbles at the centres of older SNR
as shown by Chevalier (1983a).
These bubbles may extend 10s of parsec and extend half-way to the outer edge of the SNR.
Inside the bubble, the CR energy density exceeds the thermal energy density,
and the maximum CR energy $T_{max}$ 
exceeds that of CR close to the shock.
Despite the large CR energy density in the interior, CR protons are not strong emitters of $\gamma $-rays 
because of the low interior mass density and the consequent lack of thermal protons 
as targets for proton-proton interactions (see figure 1).
CR electrons may be more detectable in the interior especially if they interact with 
a uniform photon density to emit inverse Compton radiation.
The radio synchrotron luminosity in the interior is uncertain since it depends on the
unknown magnitude of the magnetic field.
The interior magnetic field is strongly reduced
by adiabatic expansion
but magnetic field amplification at the shock
prior to expansion may compensate for this.
Given the low predicted $\gamma$-ray emission by protons and
the uncertainties in the radio emission,
inverse Compton emission from CR electrons appears to be the most accessible
signature of the presence of a CR bubble inside a blast wave.

The discussion presented in this paper may be applied to blast waves launched by
any rapid energy release such as may occur in the centre of the Galaxy
or any other galaxy, leading possibly to the formation of the 
Galactic Fermi bubbles (Carretti et al 2013)
or the SNR-like shocks observed in Centaurus A (Croston et al 2009).


\section{Conclusions}

Our principal conclusions are that:
\newline
$\bullet $ 
SNR in the Sedov phase contain a CR bubble at their centre 
that extends to a quarter or a half of the SNR radius as
previously shown by Chevalier (1983a).
\newline 
$\bullet $
Adiabatic expansion serves to increase rather than decrease the efficiency
of Sedov-phase SNR as producers of Galactic CR.
\newline 
$\bullet $
Galactic CR can be divided into two populations: 
(A) CR at higher energies that escape upstream of the shock into the ISM
as part of the acceleration process as discussed by Bell et al (2013), 
(B) CR with energies up to about 200GeV
constituting CR bubbles that are released into the Galaxy at the end of the SNR's life.
\newline 
$\bullet $
The intersection of the two populations
may tentatively be identified with the change in spectral index
detected by Adriani et al (2011) at $\sim 200$GeV.
\newline 
$\bullet $
The CR electrons in the bubble may be detected by inverse Compton $\gamma -$rays
but CR protons may be relatively undetectable due to the low mass density
in the centre of blast wave.  
Synchrotron radio emission depends upon the magnetic energy
density at the centre of a blast wave.
\newline 
$\bullet $
The above discussion may be applicable to blast waves originating from
the centres of our Galaxy or other galaxies.

\section{Acknowledgements}

I especially thank Brian Reville, Klara Schure and Gwenael Giacinti for many enlightening 
discussions relevant to this work; also Bojan Arbutina for interesting discussions
during the National Conference of Astronomers of Serbia (Sept 2014),
and an anonymous referee for helpful comments.

I thank the Aspen Center for Physics and the NSF Grant no.1066293 for hospitality during the 
workshop on
"Astrophysical mechanisms of particle acceleration and escape from the 
accelerators", Sept. 1-15, 2013, organised by Mikhail Malkov.

The research leading to these results has received funding
from the European Research Council under the European
Community's Seventh Framework Programme (FP7/2007-
2013) / ERC grant agreement no. 247039 and from grants 
ST/H001948/1 and ST/K00106X/1
made by the UK Science Technology and Facilities Council.

\section{References}

Abdo A.A. et al, 2010, Science 327, 1103
\newline
Achterberg A., 1983, A\&A, 119, 274
\newline
Adriani O., et al, 2011, Science, 332, 69
\newline
Ahn HS et al, 2010, ApJLett 714, L89
\newline
Axford W.I., Leer E. \& Skadron G., 1977, Proc 15th Int. Cosmic Ray Conf., 11, 132
\newline
Bell A.R., 1978, MNRAS, 182, 147
\newline
Bell A.R., 2004, MNRAS, 353, 550
\newline
Bell A.R., 2013, Astropart Phys 43, 56
\newline
Bell A.R., Schure K.M. \& Reville B., 2011, MNRAS, 418, 1208
\newline
Bell A.R., Schure K.M., Reville B. \& Giacinti G., 2013, MNRAS, 431, 415
\newline
Berezhko E.G., Ksenofontov L.T. \& V\"{o}lk H.J., 2003, A\&A 412, L11
\newline
Berezhko E.G \& V\"{o}lk H.J., 2004, A\&A 427, 525
\newline
Berezhko E.G. \& V\"{o}lk H.J., 2007, ApJLett 661, L75
\newline
Blandford R.D. \&  Ostriker J.P., 1978, ApJ, 221, L29
\newline
Blasi P., Amato E. \& Serpico P.D., 2012, Phys Rev Lett 109, 061101
\newline 
Carretti E. et al., 2013, Nature 493, 66
\newline
Castro D., Slane P., Patnaude D.J. \& Ellison D.C., 2011, ApJ 734, 85
\newline
Chevalier, R.A., 1982, ApJ 258, 790
\newline
Chevalier R.A., 1983a, ApJ, 272, 765
\newline
Chevalier, R.A., 1983b, Proc. 18th Int Cosmic Ray Conf ( Bangalore), 2, 314
\newline
Croston J.H. et al, 2009, MNRAS, 395, 1999
\newline
Hillas A.M., 1984, ARA\&A 22, 425
\newline
Krymsky G.F., 1977, Sov Phys Dokl, 23, 327
\newline
Kulsrud R. \& Pearce W.P., 1969, ApJ 156 445
\newline
Lagage O. \& Cesarsky C.J., 1983a, A\&A 118 223
\newline
Lagage O. \& Cesarsky C.J., 1983b, ApJ 125 249
\newline
Lerche I., 1967, ApJ 147, 689
\newline
Ostrowski M., 1999, A\&A 345, 256
\newline
Ptsukin V., Zirakashvili V. \& Seo E-S., 2013, ApJ 763, 47
\newline
Schure K.M. \& Bell A.R., 2013, MNRAS 435, 1174
\newline
Ting S., 2013, '{\it The AMS spectrometer on the Internation
Space Station}', Highlight Talk, 33rd Int Cosmic Ray Conf (Rio de Janeiro)
\newline
Tomassetti N., 2012, ApJL 752, L13
\newline
Truelove J.K. \& McKee C.F.. 1999, ApJSS 120, 299
\newline
Vink J. \& Laming J.M., 2003, ApJ, 584, 758
\newline
V\"{o}lk H.J., Berezhko E.G. \& Ksenofontov L.T., 2005, A\&A, 433, 229
\newline
Wentzel D.G., 1974, ARA\&A 12, 71
\newline
Zirakashvili V.N. \& Ptuskin V.S., 2008, ApJ 678, 939


\vskip 1 cm

\noindent
{\bf APPENDIX A:  THE SOLUTION AT SMALL RADIUS}

In this appendix we derive the asymptotic profiles close to zero radius.
Multiplying equation 3 by $R^3/r_s^3$ and differentiation with respect to radius gives
$$
R^3 \frac{\partial P}{ \partial r}+\frac{3\rho r^2}{\rho _0}P=
\left (
\frac{5}{3}(1-\epsilon) \left(\frac{\rho}{\rho _s}\right )^{2/3}+
\frac{4}{3}\epsilon \left(\frac{\rho}{\rho _s}\right )^{1/3}
\right )
\frac{1}{\rho _s}\frac{\partial \rho}{\partial r}
\eqno{(A1)}
$$
After rearrangement,
$$
\left (
3P+ \frac{3\int _0 ^r \rho r^2 dr}{\rho r^3}\ r \frac{\partial P}{\partial r}
\right )r^2
\hskip 10 cm
$$
$$
\hskip 0.3 cm=\frac {\rho _0 P_s}{\rho _s^2}
\left (
\frac{5}{3}(1-\epsilon) \left(\frac{\rho}{\rho _s}\right )^{-1/3}+
\frac{4}{3}\epsilon \left(\frac{\rho}{\rho _s}\right )^{-2/3}
\right )
\frac{\partial \rho}{\partial r}
\eqno{(A2)}
$$
From equation 5 $\partial P/\partial r\rightarrow 0$ as $r\rightarrow 0$
since $u \rightarrow 0$ as $r\rightarrow 0$
and $\partial u/\partial r$ is finite.
The pressure must be non-zero at $r=0$ since the motions are sub-sonic
at the centre of the blast wave ($u \rightarrow 0$).
Hence the term including $\partial P/\partial r$ can be neglected in equation A2.
We define $P_c$ as the pressure at $r=0$ and integrate equation A2
with respect to $r$ to obtain
$$
\frac{5}{2}(1-\epsilon)  \left(\frac{\rho}{\rho _s}\right )^{2/3}
+ 4 \epsilon  \left(\frac{\rho}{\rho _s}\right )^{1/3} 
-\frac{\rho _s P_c}{\rho _0 P_s}  \left(\frac{r}{r _s}\right )^{3}
=0
\eqno{(A3)}
$$
This quadratic in $(\rho/\rho _s)^{1/3}$ can be solved to obtain an expression for
$\rho$ which is reproduced in equation 7.
$$
\frac {\rho}{\rho _s }
=
\left \{ 
\frac{1}{5(1-\epsilon)}
\left [ \left (16 \epsilon^2 + \frac{10 (1-\epsilon) \rho _s P_c r^3}{\rho _0 P_s r_s^3} \right )^{1/2}
- 4\epsilon \right]
\right \}^3
\eqno{(A4)}
$$
At the centre ($r \rightarrow 0$) $\rho \propto r^9$ unless $\epsilon=0$
in which case  $\rho \propto r^{9/2}$.
The strong dependence of $\rho$ on $r$ strengthens the assertion above from equation 5 that 
$\partial P/\partial r$ can be neglected in equation A2.

\vskip 1 cm

\noindent
{\bf APPENDIX B:  THE MAXIMUM CR ENERGY AT A SHOCK}

Equation 15 for the maximum CR energy at a shock was derived by Bell et al (2013) on the
basis that a sufficient electric current must escape upstream of the shock to 
amplify the magnetic field through the growth of the NRH instability by
about 5 e-foldings at its maximum growth rate.  
This determines the energy of the escaping CR since for
a given CR energy flux set to a fixed fraction of $\rho u_s^3$
the electric current is too small if the energy of CR carrying the current
is very large.
Conversely, if the energy of escaping CR is too low
the CR current is large and the instability grows so rapidly
that the magnetic field is strongly amplified and the CR are unable
to escape upstream.
For more details of the model see Bell et al (2013).

The argument of Bell et al (2013) and the derivation of equation 15
for the maximum CR energy were based on the assumption that the
NRH instability is active and dominant.
This is true for young SNR with high shock velocities,
but the NRH instability is inactive for SNR in
the late Sedov phase.
When the CR current drops below 
a characteristic value 
$j_c=B/(\mu _0 r_g)$
the  ${\bf j_{CR}}\times {\bf B}$ force is too weak 
to overcome the tension in the magnetic field
and the NRH instability ceases to operate.
For a magnetic field of $5\mu{\rm G}$ and an electron density
of $1{\rm cm}^{-3}$, $j_{CR}$ drops below the crossover value
$j_c$ when the shock velocity falls below $1,000{\rm km\ s}^{-1}$
At shock velocities below this the Alfven instability 
(Lerche 1967, Kulsrud \& Pearce 1969, Wentzel 1974) driven by CR streaming
dominates.
The Alfven instability causes the growth of Alfven waves
in spatial resonance with the CR Larmor radius.
Because Alfven waves are natural modes of the system
they are undamped by tension in the magnetic field
and can grow even if the growth rate drops below the
natural frequency of the wave.  
The NRH and Alfven instabilities drive modes 
with opposite circular polarisations.
In this appendix we set out the derivation of the
maximum growth rates of both the Alfven and NRH
instabilities for monoenergetic streaming CR
using the formalism of Bell (2004),
showing that the maximum growth rates for each instability are 
given by very similar expressions, differing only by 10 percent.
The similarity of the two growth rates was previously
noted by Zirakashvili \& Ptuskin (2008).

Because of the similar growth rates
the estimate of the maximum CR energy
based on instability growth rates derived by Bell et al (2013)
for the NRH instability 
at high shock velocities also applies
to the Alfven instability 
at low shock velocities.
Crucially for this paper, equation 15
can be applied to SNR throughout the Sedov phase.

The dispersion relation for CR-driven instability can be found in 
equation 7 of Bell (2004) (see also Achterberg (1983)).
It includes both the Alfven and NRH instabilities.
The dispersion relation is
$$
\omega ^2 =k^2 v_A^2 + (1-\sigma) \frac {k j_{CR} B_{||}}{\rho}
\eqno{(B1)}
$$
where $k$, $j_{CR}$ (named $j_{||}$ in Bell (2004)) and $B_{||}$ are the wavenumber,
CR electric current and magnetic field respectively, each aligned parallel
to the shock normal. $v_A$ is the Alfven speed.
A small term in $\omega /k u_s$ has been omitted from equation 7 of Bell (2004)
as justified therein.
The function $\sigma$ describes the response of the streaming CR
to perturbations in the magnetic field.
For monoenergetic CR with a Larmor radius $r_g$
propagating diffusively relative to the background plasma, and $\lambda=1/kr_g$,
$$
\sigma = \frac{3}{4}\lambda (1-\lambda ^2)
\left [
\ln \left ( \frac{\lambda +1}{\lambda-1} \right ) \right ]
+\frac{3}{2}\lambda ^2
\eqno{(B2)}
$$
for long wavelengths, $k<r_g^{-1}$, $\lambda >1$, and 
$$
\sigma = \frac{3}{4}\lambda (1-\lambda ^2)
\left [
\ln \left (\frac {1+\lambda }{1- \lambda} \right ) +i\pi \right ]
+\frac{3}{2}\lambda ^2
\eqno{(B3)}
$$
for short wavelengths, $k>r_g^{-1}$, $\lambda <1$.
The imaginary term ($i \pi$) at short wavelengths
results from the spatial resonance
with the CR Larmor radius.
No such resonance occurs at wavelengths longer than
the CR Larmor radius,
which accounts for the absence of the imaginary term for $kr_g<1$.

When the CR current $j_{CR}$ is small and magnetic perturbations
grow by the Alfven instability,
$k^2 v_A^2$ 
dominates the real part of the right hand side of equation B1.
In this limit, and with $kr_g>1$,
$$
\omega ^2 =k^2 v_A^2 - \frac{3 \pi i}{4} 
\left (
\frac {k^2 r_g^2 -1}{k^3r_g^3}
\right )
\frac {k j_{CR} B_{||}}{\rho}
\eqno{(B4)}
$$
In the limit of small $j_{CR}$,
the maximum growth rate is
$$
\gamma_{max}
=
\frac{ \pi  }{4}
\sqrt{\frac{\mu _0}{3\rho}}j_{CR}
=
0.45
\sqrt{\frac{\mu _0}{\rho}}j_{CR}
\eqno{(B5)}
$$
which occurs when $kr_g=\sqrt{3}$.

In contrast, when 
the CR current $j_{CR}$ is large and magnetic perturbations
grow by the NRH instability,
the maximum growth rate occurs at wavelengths much shorter
than the CR Larmor radius ($\lambda \ll 1$). 
Both the real and imaginary parts of $\sigma \ll 1$
can then be neglected giving
$$
\omega ^2 =k^2 v_A^2 + \frac {k j_{CR} B_{||}}{\rho}
\eqno{(B6)}
$$
In the appropriate polarisation, $k j_{CR} B_{||}<0$,
$$
\omega  = \pm i \left (  \frac {|k j_{CR} B_{||}|}{\rho}
-k^2 v_A^2
\right )^{1/2}
\eqno{(B7)}
$$
and the maximum growth rate is
$$
\gamma_{max}
=
0.5
\sqrt{\frac{\mu _0}{\rho}}j_{CR}
\eqno{(B8)}
$$
which occurs when $|k|=0.5 \mu_0 j_{CR}/B_{||}$.

\begin{figure}
\includegraphics[angle=0,width=8cm]{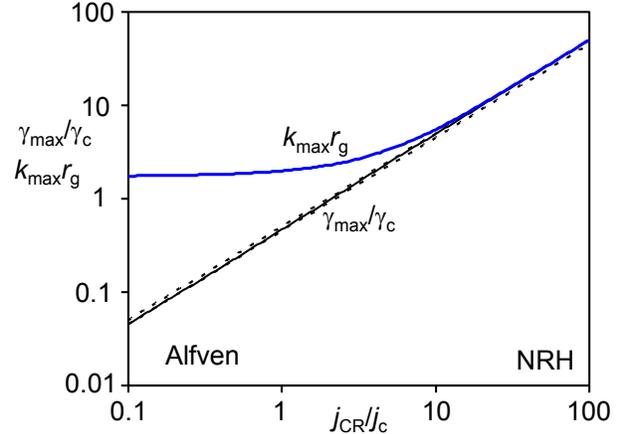}
\caption{
Plots of the maximum growth rate $\gamma _{max}$
and the wavenumber $k_{max}$ at which the growth rate is maximum
against the CR current $j_{CR}$.
$k_{max}$ is normalised to the CR Larmor radius $r_g$.
$\gamma _{max}$ is normalised to $\gamma_c =  v_A/r_g$.
$j_{CR}$ is normalised to $j_c =B/\mu _0 r_g$.
The transition from the Alfven regime to the NRH regime occurs
where $j_{CR}=2 \pi j_c$, ie when $j_{CR}=k_gB/\mu _0$ where $k_g=2\pi/r_g$.
The dotted lines correspond to the asymptotic limits
$\gamma_{max}=0.45 \gamma_c (j_{CR}/j_c)$ and $\gamma_{max}=0.5 \gamma_c (j_{CR}/j_c)$
for the Alfven and NRH instabilities respectively.
}
\label{fig:figure8}
\end{figure}

Although the NRH and Alfven instabilities operate 
in different ways and in different polarisations, equations B5 and B8
show that the maximum growth is very similar
in the low $j_{CR}$ Alfven limit and the
high $j_{CR}$ NRH limit.
The maximum growth rate for $j_{CR}$ across the range
from the Alfven to the NRH limit
is plotted in figure 8.
To a good approximation the maximum growth rate
can be assumed to be 
$0.5 \sqrt{({\mu _0}/{\rho})} j_{CR}$
across the whole range of $j_{CR}$. 
Consequently,
equation 15 provides a good estimate of the maximum CR energy at 
a shock at all times during the Sedov phase of SNR expansion.

The discussion in this appendix has treated the CR distribution as monoenergetic.
This is reasonable for escaping CR which have to reach a certain energy before they
escape and are not accelerated beyond this energy.
Bell (2004) derives the dispersion relation
for a $T^{-2}$ CR distribution and similar results can be 
obtained from the plots of the real and imaginary parts of $\sigma$
in figure 1 of that paper. 

\end{document}